\documentstyle[pre,aps,epsfig,preprint]{revtex}
\begin{document}

\title{Secure Digital Signal Transmission 
by Multistep Parameter Modulation and Alternative Driving of 
Transmitter Variables}

\author{P. Palaniyandi and M. Lakshmanan}
\address{Centre for Nonlinear Dynamics, Department of Physics \\
Bharathidasan University, Tiruchirapalli - 620 024, India.}
\date{\today}
\maketitle

\begin{abstract}

The idea of secure communication of digital signals via chaos
synchronization has been plagued by the possibility of attractor
reconstruction by eavesdroppers as pointed out by  P\'erez and
Cerdeira.  In this Letter, we wish to present a very simple mechanism
by which this problem can be overcome, wherein the signal is
transmitted via a multistep parameter modulation combined with
alternative driving of different transmitter variables, which makes the
attractor reconstruction impossible.  The method is illustrated by
means of the Lorenz system and Chua's circuit as examples.

\end{abstract}


The work of  Pecora and Carroll [1990,1991] on synchronization of
chaotic signals has suggested the possibility of secure communication
using chaos synchronization. They have shown that two chaotic systems
can be synchronized when they are linked by a common signal (driving
signal) provided the Lyapunov exponents of the subsystem are all
negative.  Following this suggestion a number of methods have been
proposed for secure communication [Lakshmanan \& Murali, 1996; Hayes et
al., 1993; Cuomo \& Oppenheim, 1993; Murali \& Lakshmanan, 1993a,1993b;
Kocarev \& Parlitz, 1995; Zhang et al., 1998] with the help of chaotic
signals.  Cuomo and Oppenheim [1993] have shown that a small difference
between the corresponding parameters in the drive and response systems
will cause synchronization frustration between the transmitter and
receiver variables.  Using this property, they have suggested a very
simple method for digital signal transmission and illustrated it with
the Lorenz system.  However, P\'erez and Cerdeira [1995] have pointed
out that it is possible to reconstruct the message by an eavesdropper
from a simple return map formed by the extrema of the modulated driving
signal (Here, the driving signal is used as a carrier signal in the
digital signal transmission. In our method the digital message is
imposed on this carrier signal through parameter variation. So, we call
the driving signal bearing the digital message as modulated driving
signal and the modulation as parameter modulation.), even without
synchronization with the receiver system.  To overcome the above
problem several methods have been suggested by many authors recently
[Murali \& Lakshmanan, 1998; Mensour \& Longtin, 1998; Minai \&
Pandian, 1998]:   Private communication using compound chaotic signal
technique, using delay-differential equations technique, communication
through noise and so on.  However, most of the above methods are very
complicated and are often difficult to implement.  In this Letter, we
describe a simple method by which the digital signal can be highly
masked with the chaotic signal by using a multistep parameter
modulation and show how it can be further complicated by introducing
two different drive signals alternatively instead of a single drive
signal. As a result, reconstruction of the message becomes virtually
impossible and this provides further development in the technique of
secure communication using chaotic signals.

In order to present our method, we make use of the dynamical system
used by Cuomo and Oppenheim [1993] namely the Lorenz system. The
associated evolution
equations at the transmitter end are
\begin{eqnarray}
\frac{dx_s}{dt} &=& \sigma (y_s-x_s),\nonumber \\ 
\frac{dy_s}{dt} &=& rx_s-y_s-x_sz_s, \\ 
\frac{dz_s}{dt} &=& x_sy_s-bz_s,  \nonumber
\end{eqnarray}
while at the receiver end, corresponding to an $x$-drive, they read as
\begin{eqnarray}
\frac{dx_r}{dt} &=& \sigma (y_r-x_r), \nonumber \\
\frac{dy_r}{dt} &=& rx_s-y_r-x_sz_r, \\ 
\frac{dz_r}{dt} &=& x_sy_r-bz_r, \nonumber
\end{eqnarray}
where $\sigma=16.0, r=45.6$ and  the parameter $b$ chosen for the
purpose of modulation can have  either the value $4.0$ or $4.4$.  The
numerical value of the parameter  $b$ at the receiver end is kept at a
constant value 4.0, while in the transmitter circuit it is changed
between two values, namely 4.0 and 4.4, when the digital signal is zero
and one, respectively. The transmitted digital message is reconstructed
by using the synchronization error power $(x_r-x_s)^2$.  It is
negligible when the transmitter and the receiver systems are
synchronized, while it has some finite value when they are not
synchronized. Thus the receiver will synchronize with the transmitter
when the parameter $b$ in the latter takes a value $4.0$, while
asynchronization takes place when $b=4.4$ at the transmitter. It may be
noted that instead of the $x$-drive at the receiver end one can as well
use the $y$-drive in which case the eq.(2) will be modified as
\begin{eqnarray}
\frac{dx_r}{dt} &=& \sigma (y_s-x_r),\nonumber \\
\frac{dy_r}{dt} &=& rx_r-y_r-x_rz_r, \\ 
\frac{dz_r}{dt} &=& x_ry_s-bz_r. \nonumber
\end{eqnarray}

In the work of P\'erez and Cerdeira [1995],  it has been shown that even
without any receiver circuit, one can unmask the message from the
modulated drive signal.  In this reference, the maxima $X_m$ and minima
$Y_m$ are collected from the modulated driving signal and the new
varibles $A_m=(X_m+Y_m)/2$ and $B_m=X_m-Y_m$ are defined. The plot of
the return map for $A_m$ Vs $B_m$ is shown in fig.  1.  Note that there
are 3 segments in the attractor each one of which is further split into
two strips (one corresponding  to the high state and the other
corresponding to low state of the digital message). It is obvious that
the split in the attractor is due to the change of the parameter $b$ at
the sender end between the values $4.0$ and $4.4$.  From the return map
one can easily unmask the message by finding the strip  on which the
point$(A_m,B_m)$ falls in each segment of the attractor.  We can call
the above procedure as a single-step parameter modulation, since $b$
can have only one value for one state of the message (that is,  $b$ can
take only a single value $4.4$ for any high state in the digital
message and only $4.0$ for the low state digital signal at the
transmitter).  The crux of the problem is then that from the nature of
the chaotic attractor and the return map of the Lorenz system, one can
have only eight possibilities of message reconstruction from which one
can easily discover the actual message.  Then, one possible way to
overcome such a reconstruction is to increase the number of possible
ways by which message can be reconstructed to an unmanageable level and
thereby eliminating the possibility of identifying the correct
message.

In our work, we have effected two changes in the mode of transmission
of the digital message by the parameter modulation in order to
complicate the attractor in the return map.  Firstly, in the sender
part, instead of using a single-step parameter modulation, we use a
multi-step parameter modulation, where for one state of the digital
message (either $1$ or $0$), the parameter  $b$ in the sender equation
is assigned to have any one of the '$n$' preassigned values ($n>1$ and
sufficiently large), while at the receiver part we use $n$ receiver
subsystems with different parameters.  We call the number $n$ as the step
of the modulation.  Secondly, we will use both $x$ and $y$ signals to
drive the receiver subsystems alternatively during the transmission as
indicated in eqs.(2) and (3).

The block diagram of the transmitter and the receiver with modulation
step $n$ is shown in fig. 2, where R$_1$,R$_2$,...R$_{n}$ are the
subsystems of the receiver.  Each one of the $R_i'$s is assigned a
specific value among the $n$ chosen parameter values which are used to
impose the  high state of the digital message on the driving signal.
The synchronization error power at the output of each subsystem is fed
into low pass filters(LPFs) separately.  Then the filtered signal is
converted into a digital signal at threshold detectors(TD)[Cuomo \&
Oppenheim, 1993], which is then passed through NOT gates in order to
invert and to confirm with the original assignment of high and low
states of the digital signal.  An OR gate at the receiver combines all
the outputs from the NOT gates.  Thus, when the digital message imposed
on the driving signal is '$1$', any one of the $n$ subsystems at the
receiver will synchronize and the output at the OR gate will be '$1$'.
None of the subsystems will synchronize, if the modulated digital
message  is zero, and hence '$0$' will be the output of the OR gate.

We  illustrate our method in the case of the Lorenz system with the
same parameters used by P\'erez and Cerdeira [1995] and with the
modulation step as small as $n=5$.  The Lorenz system (1) is known to
exibit chaotic behavior when the parameter  $b$ takes any value
between  $1.5$ and $6.8$.  One can assign a specific value to $b$
within this range for the purpose of modulation.  We select the region
$b=\{3.0, 4.0\}$ for the illustration of our method.  As an example,
the parameter $b$ in the transmitting system is allowed to take any one
of the five values $3.1, 3.3, 3.5, 3.7, 3.9$ for high state of the
digital signal, while it can have any one the values $3.2, 3.4, 3.6,
3.8, 4.0$ for the low state.  In the receiver part, we use 5 subsystems
with the modulation parameter $b$ fixed at $3.1, 3.3, 3.5, 3.7, 3.9$
respectively.

The method works as follows. Suppose we have a high state in the
digital message, to start with. Then, the receiver is driven by the
modulated driving signal with modulation parameter $b=3.1$.  For the
next high state in the message the modulation is done with $b$ = $3.3$
and this process continues upto the $b$ value $3.9$.  Then the value of
$b$ is reset to $3.1$.  This procedure is also applicable to the low
state of the message  but with the modulation parameter($b$) taking the
values $3.2, 3.4, 3.6, 3.8, 4.0$ in that order.   In the receiver
part,  all the subsystems are driven by the modulated driving signal.
However, the subsystems  $R_1,R_2,R_3,R_4$ and $R_5$ will synchronize
only for the  values of modulation parameter at the transmitter, $3.1,
3.3, 3.5, 3.7, 3.9$ respectively.  When the synchronization is achieved
in a subsystem, the error power will be below the threshold level and
so the message obtained from TD will be at the low state. It is
inverted to the high state using the NOT gate.  But a low state is
attained by the digital message at the NOT gate, if the subsystem is
not synchronized.  Thus a high state message is obtained as an output
at the OR gate if any one of the subsystems is synchronized and a low
state digital message is recovered when asynchronization takes place at
all the subsystems. A test digital message transmitted and recovered by
this scheme by numerical simulation is shown in fig. 3.

Actually the above possibility is only a test case and has been
discussed for illustrative purpose.  One can modulate $b$ in a random
way also (that is,  at the transmitter circuit, $b$ can be assigned any
one of the $n$ chosen values in a random manner) rather than in a
predetermined way.  The method presented here is applicable equally
well to such possibilities also.

The return map constructed using the modulated  driving signal is shown
in fig. 4. Here we  have 10  strips in each segment.  This corresponds
to $2 \times n$ strips, where $n$ is the step of the parameter
modulation.  Because of the complicated nature of the return map it is
very difficult to find the strip  on which the point$(A_m,B_m)$ falls
in each segment of the attractor and so it is difficult to unmask the
message from the return map. Even if one can identify the location of
the points in the strip, still one has a serious difficulty in
assigning '0' and '1' to the strips since there are $(2^{2n}-2)^3-1$
chances to make a mistake,  for a successful reconstruction of the
message.  For $n=1$, there are 7 chances [P\'ere \& Cerdeira, 1995]  of
making a mistake during the unmasking of the message from the modulated
driving signal and for $n=5$, this chances increases to a high value of
1,067,462,647 (of the order of $10^9$).  So it is almost impossible to
unmask the message masked by our scheme, even at as low value as
$n=5$.

We can complicate the attractor in the return map further by  the
second method.  In the Lorenz system, synchronization [Pecora \&
Carroll, 1991] is possible by both $x$ driving and $y$ driving, since
the conditional Lyapunov exponents are all negative in both the cases.
We use this property in the second method.  Here the principle of
transmitting and receiving the message is the same as described above,
but instead of using a single driving signal $x(t)$, we now use both
$x(t)$ and $y(t)$ as the driving signals alternatively. For the odd
digital bits, we drive the receiver by modulated $x(t)$ whereas the
even digital bits are transmitted using modulated $y(t)$ as the driving
signal.  The block diagram of the transmitting and receiving   systems
with modulation step $n$ is shown in fig. 5. Here we use $2n$
subsystems in the receiver, $2$ subsystems for each value of modulation
parameter (which are used for the transmission of high state of the
digital message), one with modulated $x$ driving and the other with
modulated $y$ driving. The recovered message is similar to the one
obtained in our previous method.  The return map constructed from the
modulated drive signals (modulated x(t) and modulated y(t)) for the
modulation step $n$=5 is shown in fig. 6.  From the return map, it is
impossible to unmask the message since the attractor of the $y$ drive
signal is merged  with the attractor of the $x$ driving  signal.

We have also applied our scheme to the simplest autonomous Chua circuit
generator of the chaotic signal [Madan, 1993; Chua et al., 1992].  We
have used the unnormalized circuit equations
\begin{eqnarray}
C_1 \frac{dv_1}{dt} &=& (1/R)(v_2-v_1)-f(v_1), \nonumber \\ 
C_2 \frac{dv_2}{dt} &=&(1/R)(v_1-v_2)+i_L, \\ 
L \frac{di_L}{dt} &=& -v_2,\nonumber 
\end{eqnarray} 
where 
$f(v_1)=G_bv_1+0.5(G_a-G_b)[|v_1+B_p|-|v_1-B_p|]$
during our verification with $R$ as the modulation parameter. The other
parameters value used are $C_1=10.0 \; nF$, $C_2=100.0 \; nF$ and
$L=18.0 \; mH$. The Chua circuit can also be synchronized either with
the driving signal $v_1$ or with $v_2$ and so we can again use them as
alternative drive variables. One can assign any value to the
parameter $R$ between $1555 \; \Omega$ to $1960 \; \Omega$, except in some
small regions where periodic windows occur.  Here the largest Lyapunov 
exponent calculated has  positive values and hence the circuit exibits
chaotic behaviour. We find from our numerical studies that it is even
more difficult to extract message in this case with multiparameter  
modulation compared to the Lorenz system, due to the nature of the
chaotic attractor.

Thus, in this Letter, we have described two simple and straightforward
methods to complicate the attractor in the return map of  the modulated
driving signal: One with multi-step parameter modulation by using a
number of values to the modulation parameter $b$ and the other by using
alternative $x$ and $y$ driving combined with multistep parameter
modulation.  We have illustrated them with the Lorenz system and Chua's
circuit as examples. We conclude that it is almost impossible to
reconstruct the digital message transmitted in our scheme by the use of
the return map formed by the modulated driving signal.

\acknowledgements

This work has been supported by the National Board of Higher
Mathematics, Department of Atomic Energy, Government of India and the
Department of Science and Technology, Government of India through the
research projects. The authors thank Dr. K. Murali for many valuable
suggestions.

\section*{REFERENCES} 

\begin{description}

\item Chua, L.O., Kocarev, Lj., Eckert, K. \& Itoh, M. [1992] ``
Experimental chaos synchronization in Chua's circuit" Int. J.
Bifurcation and Chaos {\bf 2} 705-708.

\item Cuomo, K.M. \& Oppenheim, A. V. [1993] ``Circuit implementation of
synchronized chaos with application to communication" Phys. Rev.
Lett.  {\bf 71} 65-68.

\item Hayes, S., Grebogi, C. \& Ott, E. [1993] ``Communicating with
chaos" Phys.  Rev. Lett. {\bf 70}, 3031-3035.

\item Kocarev, L. \& Parlitz, U. [1995] ``General approach for chaotic
synchronization with applications to communication"  Phys. Rev. Lett.
{\bf 74} 5028-5031.

\item Lakshmanan, M. \& Murali, K. [1996] Chaos in Nonlinear
Oscillators: Controlling and Synchronization (World Scientific,
Singapore).

\item Madan, R. N., [1993] Chua's circuit: A Paradigm for Chaos
(World Scientific, Singapore).

\item Mensour, B. \& Longtin, A. [1998] ``Synchronization of
delay-differential equations with application to private communication"
Phys. Lett. {\bf A244}, 59-70.

\item Minai, A. A. \& Pandian, T. D. [1998] ``Communicating with
noise:  How chaos and noise combine to generate secure encryption keys"
Chaos {\bf 8} 621-628.

\item Murali, K. \& Lakshmanan, M. [1993] ``Transmission of signals by
synchronization in a chaotic van der pol-duffing oscillator" Phys. Rev.
{\bf E48}, R1624-1626  

\item Murali, K. \& Lakshmanan, M. [1993] ``Synchroning chaos in driven
chua's circuit" Int. J. Bifurcations Chaos {\bf 3}, 1057-1066.

\item Murali, K. \& Lakshmanan, M. [1998] ``Secure communication using
a compound signal from generalized synchronizable chaotic systems"
Phys.  Lett. {\bf A241}, 303-310.

\item Pecora, L. M. \& Carroll, T. L. $[1990]$ ``Synchronization in
chaotic systems" Phys. Rev. Lett. {\bf 64}, 821-823 (1990);

\item Pecora, L. M. \& Carroll, T. L. $[1991]$ ``Driving systems with
chaotic signals" Phys. Rev. {\bf A44}, 2374-2383.

\item P\'erez, G. \& Cerdeira, H.A. [1995] ``Extracting messages masked
by chaos" Phys. Rev. Lett. {\bf 74}, 1970-1973.

\item Zhang, Y., Dai, M., Hua, Y., Ni, W, \& Du, G. [1998] ``Digital
communication by active-passive-decomposition synchronization in
hyperchaotic systems" Phys. Rev. {\bf E58} 3022-3027.

\end{description}

\begin{figure}
\begin{center}
\epsfig{figure=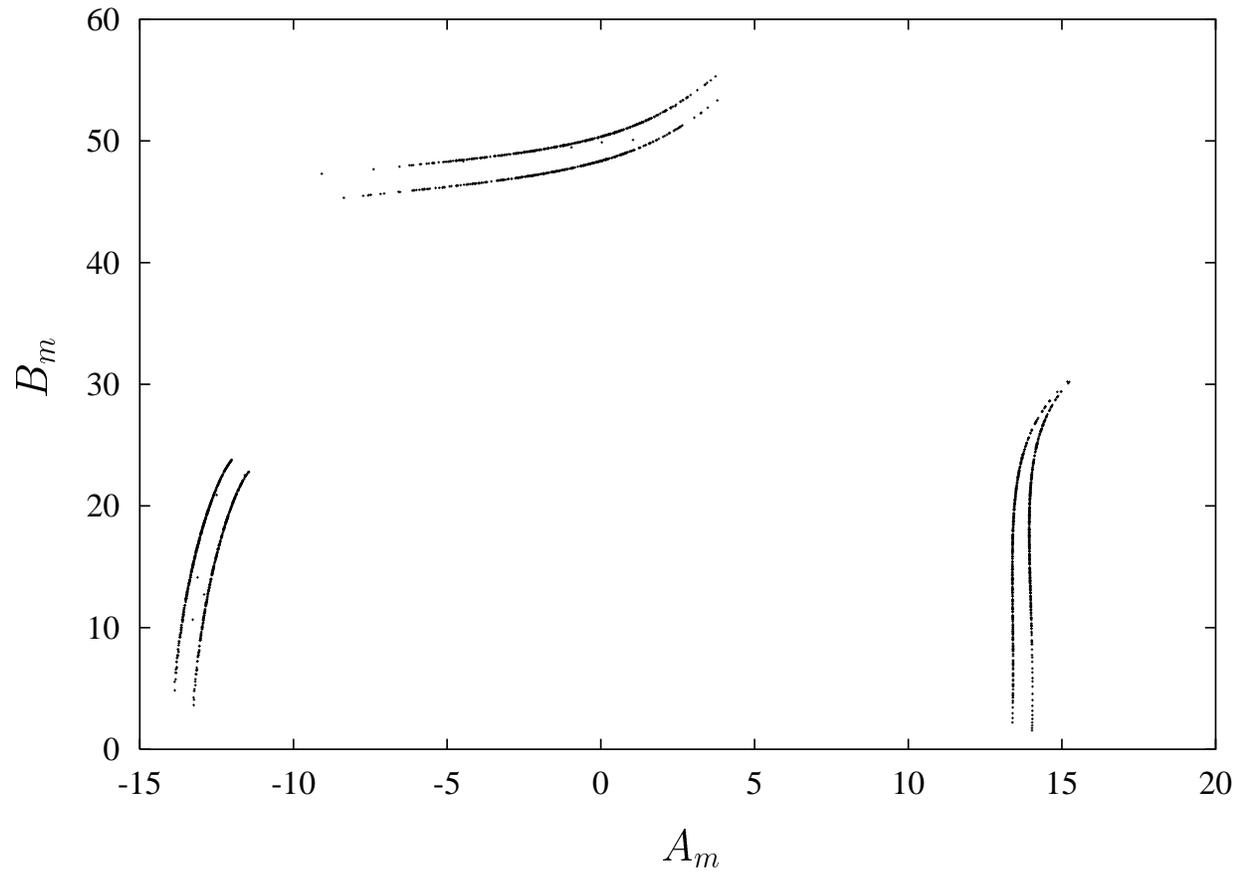, width=\columnwidth}
\end{center} 
\caption{Return map between $A_m$ and $B_m$ in single-step parameter 
modulation for the Lorenz system.}
\end{figure}

\begin{figure}
\begin{center}
\epsfig{figure=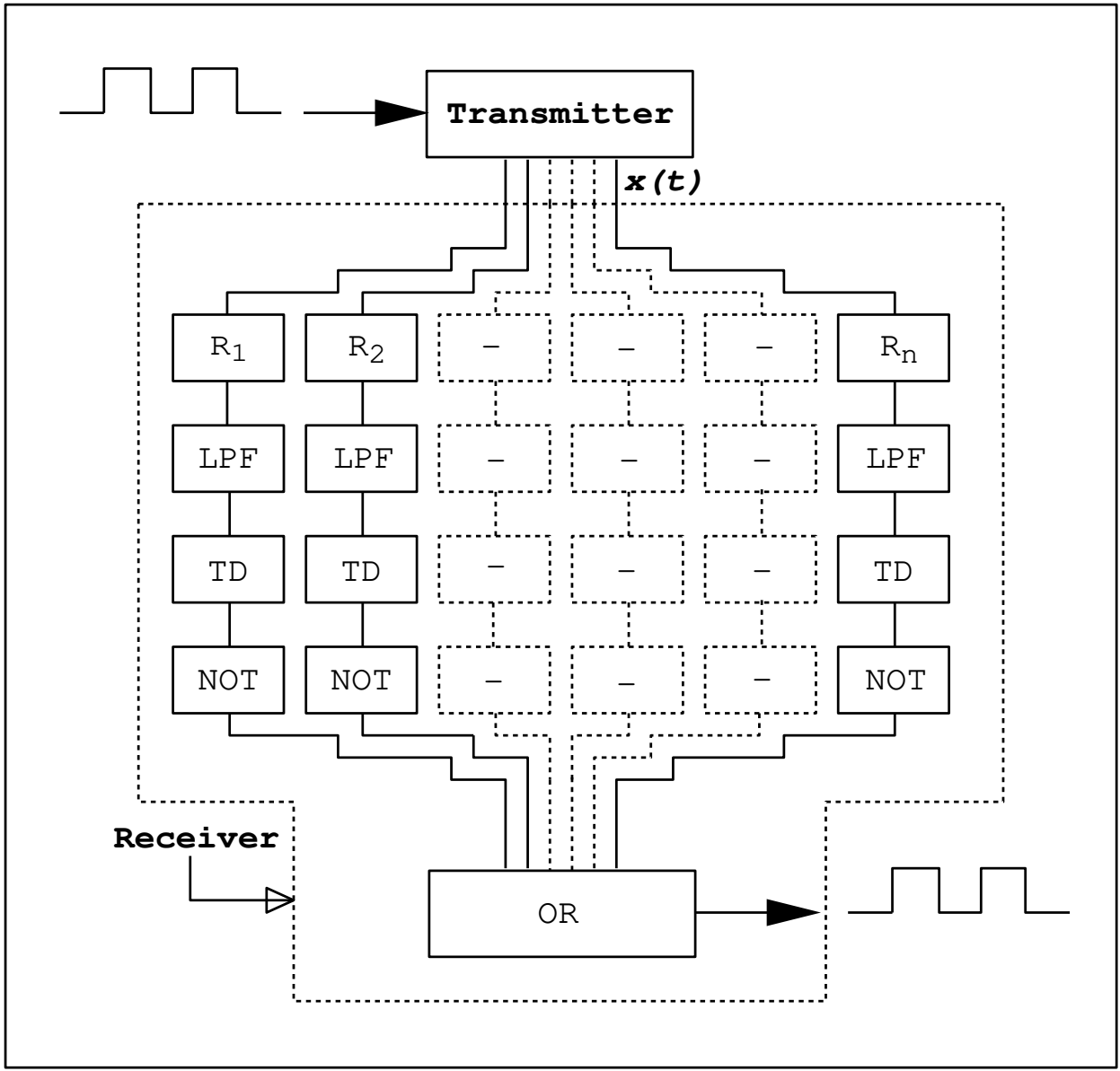, width=\columnwidth}
\end{center}
\caption{The block diagram of transmitter and receiver in multi-step
parameter modulation with $x$ driving (LPF-Low pass filter,
TD-Threshold detector).} 
\end{figure}

\begin{figure} 
\begin{center} 
\epsfig{figure=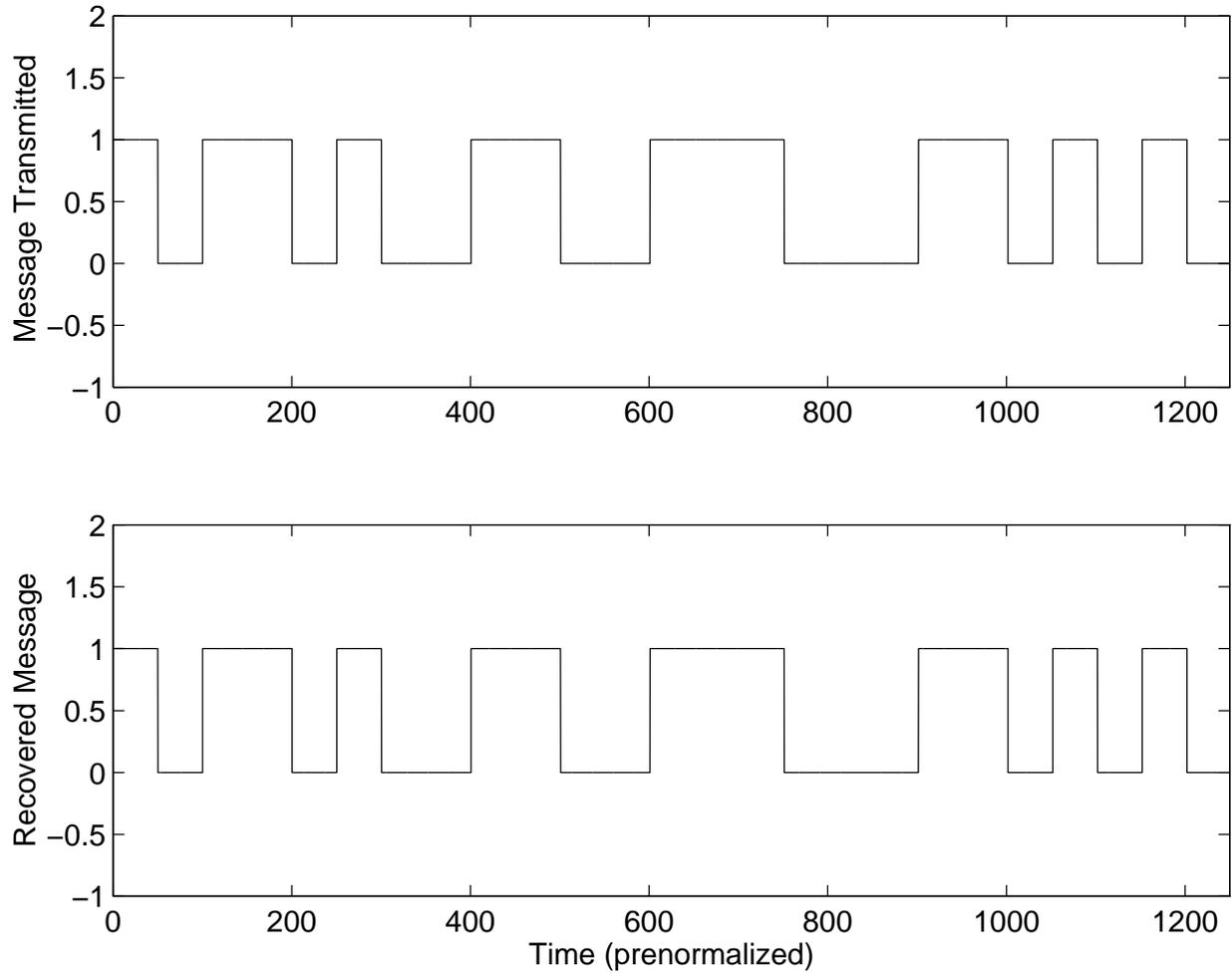,width=\columnwidth} 
\end{center} 
\caption{Transmitted and  recovered message in multi-step 
parameter modulation$(n=5)$ with $x$ driving.} 
\end{figure}

\begin{figure} 
\begin{center}
\epsfig{figure=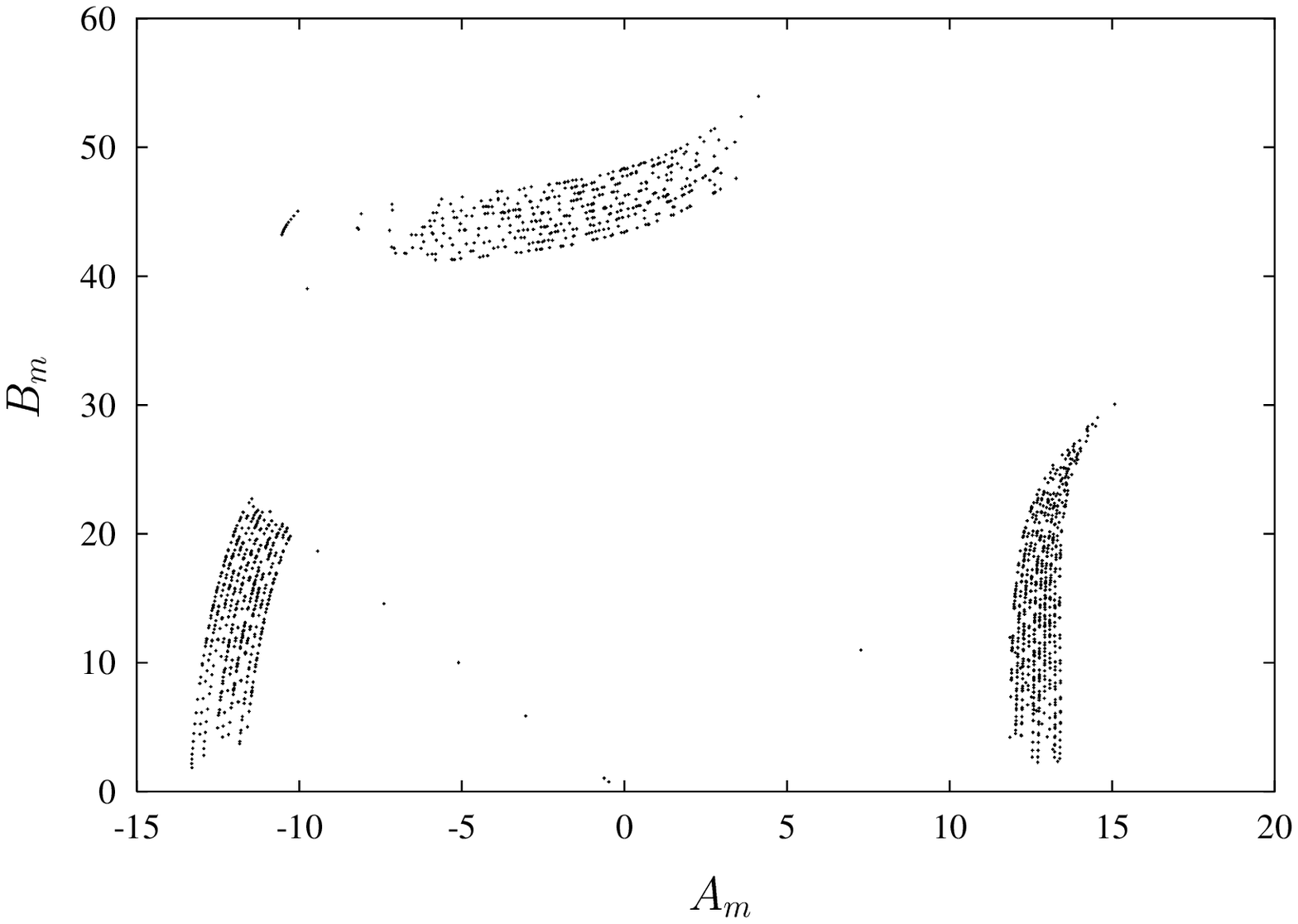, width=\columnwidth} 
\end{center}
\caption{Return map between $A_m$ and $B_m$ in multi-step parameter
modulation$(n=5)$ with $x$ driving for the Lorenz system.} 
\end{figure}

\begin{figure}[h]
\begin{center}
\epsfig{figure=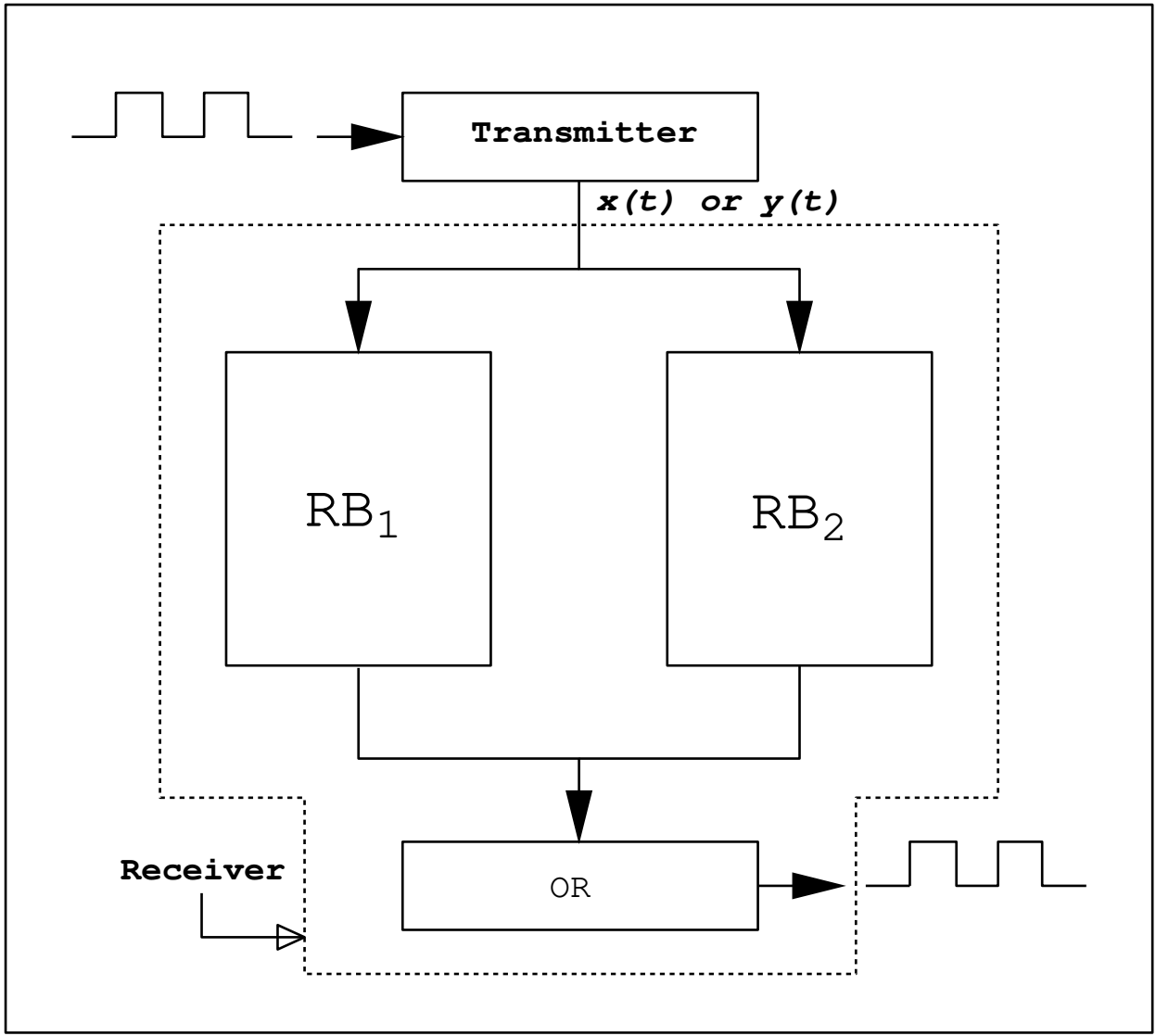, width=\columnwidth}
\end{center}
\caption{The block diagram of transmitter and receiver in multi-step
parameter modulation with $x$ and $y$ driving ($RB_1$-receiver block in
fig.2 with $x$ driving subsystems, $RB_2$-receiver block in fig.2 with
$y$ driving subsystems).}
\end{figure}

\begin{figure}
\begin{center}
\epsfig{figure=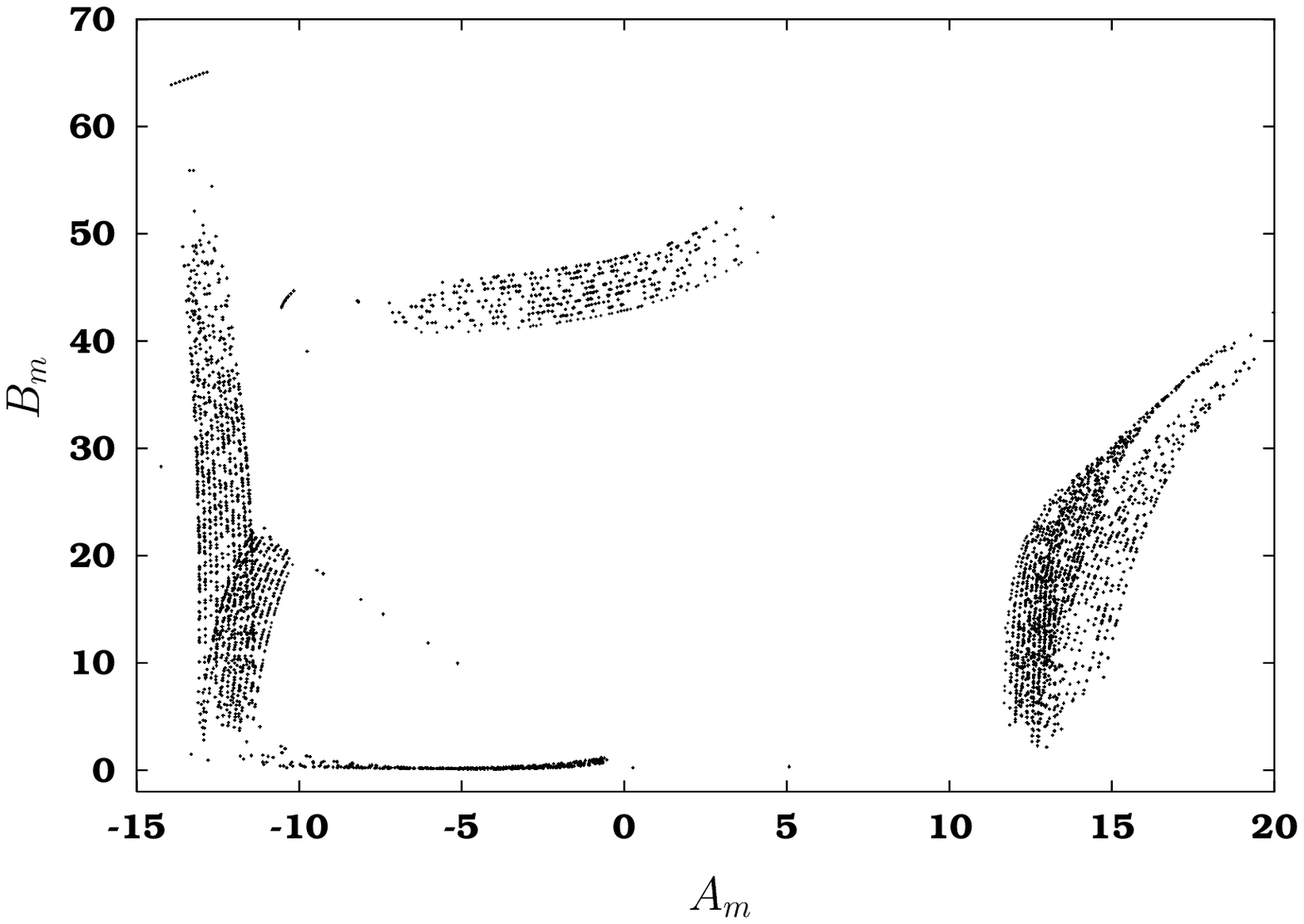, width=\columnwidth}
\end{center}
\caption{Return map between $A_m$ and $B_m$ in multi-step parameter
modulation$(n=5)$ with alternate $x$ and $y$ driving for the Lorenz
system.} 
\end{figure}

\end{document}